\documentclass[traditabstract]{aa}
\usepackage{txfonts}
\usepackage{amssymb}
\usepackage{natbib}
\usepackage{graphics}
\def\maxi{MAXI\,J1828-249}
\def\inte{{\em INTEGRAL}}

\def\swift{{\em Swift}}

\begin{document}

\title{INTEGRAL and Swift observations of the hard X-ray transient MAXI\,J1828-249}

\author{E. Filippova\inst{1,2}
\and E. Bozzo\inst{1} 
\and C. Ferrigno\inst{1}} 

\institute{ISDC, University of Geneva, Ch. d'Ecogia 16, 1290 Versoix, Switzerland
  \and Space Research Institute, Moscow, 117997, Russia; \email{Ekaterina.Filippova@unige.ch}} 

\date{Received / Accepted}

\abstract{In this paper we report on the observations performed with \inte\ and \swift\ of the first outburst detected from the hard X-ray transient 
\maxi.\ During the first about two days of the outburst, the source was observed by MAXI to undergo a very rapid transition  
from a hard to  a softer spectral state. While the hard state was not efficiently monitored because the transition occurred so rapidly, the evolution of the 
source outburst in the softer state was covered quasi-simultaneously in a broad energy range (0.6-150~keV) by the instruments on-board \inte\ and \swift.\
During these observations, the spectra measured from the source displayed both a prominent thermal emission with temperature kT$\sim$0.7~keV and a 
power-law hard component with a photon index $\Gamma \sim$2.2 extending to 200~keV. The properties of the source in the X-ray domain are 
reminiscent of those displayed by black hole transients during the soft intermediate state, which supports the association of \maxi\ 
with this class of objects. } 

 
\maketitle 

\section{Introduction}
\label{sec:intro} 

\maxi\ was discovered in outburst on 15 October 2013 (56580.91 MJD) by MAXI/GSC \citep{nakahira2013}. 
Shortly after the discovery, the source was detected at higher energies by the hard X-ray imager ISGRI on-board INTEGRAL \citep{filippova2013}. 
At discovery the flux recorded from the source was 93$\pm$9~mCrab in the 4-10~keV 
energy band and 45$\pm$2~mCrab (48$\pm$2~mCrab) in the 20-40~keV (40-80~keV) energy band. 
Follow-up observations carried out on 56581.6 MJD with the narrow field instrument on-board \swift/XRT provided the best measured position of the source at 
RA=277.2427, Dec=-25.0304 (J2000) with an estimated uncertainty of 3.6~arcsec at 90 c.l. \citep{kennea2013}. The improved position permitted us to identify 
the UV and IR counterparts of the source. The estimated magnitude of the UV source was 18.64$\pm$0.04 (stat)$\pm$0.03 (sys)
 \citep[in AB system, without correction for interstellar reddening;][]{kennea2013a}. The AB magnitudes of the optical/IR source were
$g' = 17.2\pm0.1, r' = 16.9\pm0.1, i' = 16.9\pm0.1, z' = 16.8\pm0.1, J = 16.8\pm0.1, H = 16.9\pm 0.1, K = 17.2\pm0.2$ (not corrected for the interstellar reddening). The derived spectrum of optical/IR source might be interpreted as arising because of an accretion disk \citep{rau2013}.  
Radio observations carried out about two days after the onset of the outburst did not reveal significant emission from the source \citep[the 3~$\sigma$ 
upper limit on its flux was 75~$\mu$Jy at 5.5~GHz and 57~$\mu$Jy at 9.0~GHz;][]{millerjones2013}.

In this paper we report on all \inte\ and Swift data available to our group collected during the outburst of \maxi\ from 56580 to 56607~MJD.

\begin{figure}
\centerline{\includegraphics[scale=0.3]{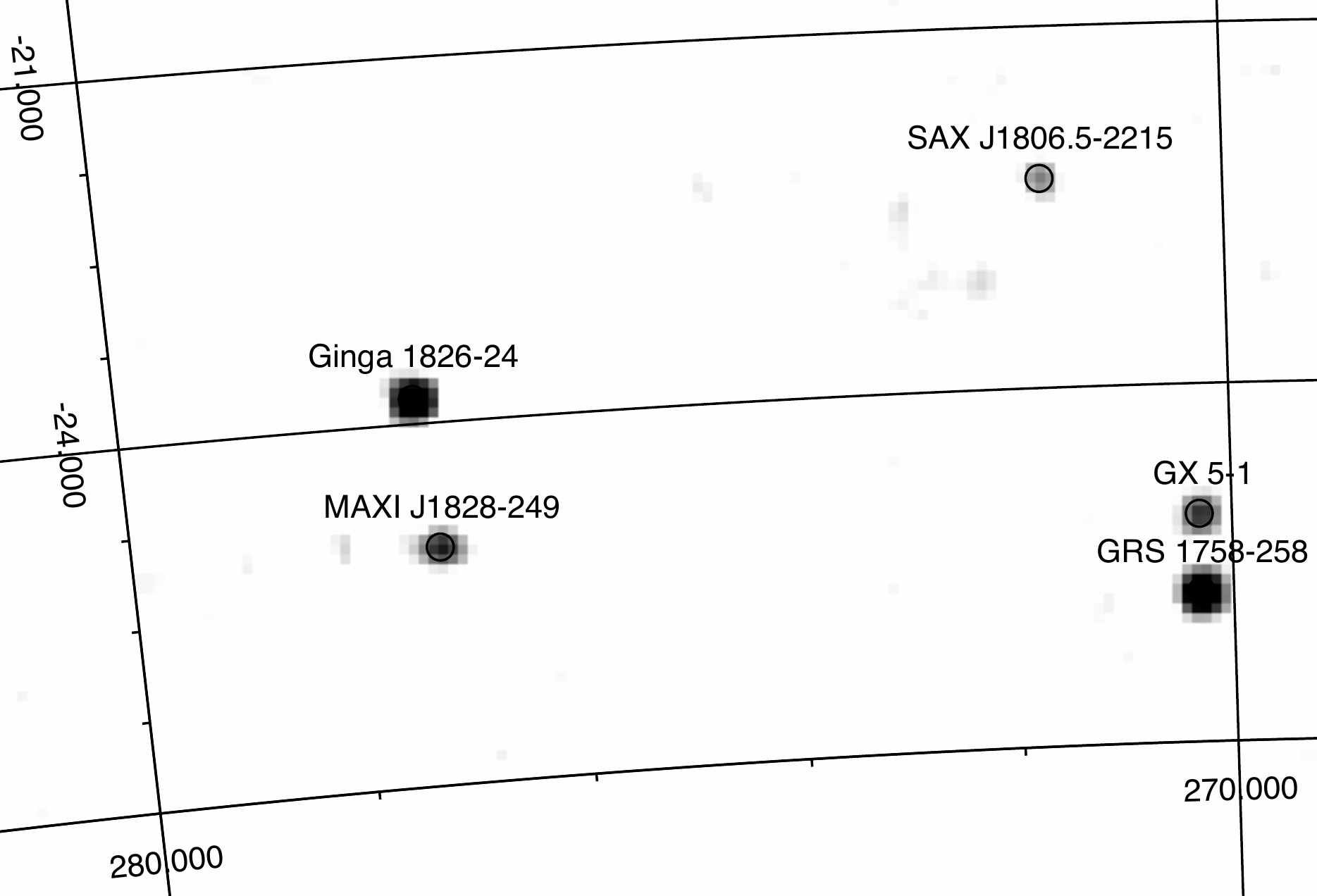}}
\caption{ISGRI mosaic of the FOV around \maxi\ (20-100 keV). 
The mosaic has been obtained from data collected during revolution 1344.} 
\label{fig:ima}
\end{figure}

\section{INTEGRAL data}

MAXI\,J1828-249 was observed by IBIS/ISGRI \citep{lebrun03,ubertini03} from  56580~MJD to 56594~MJD, that is during satellite 
revolutions 1344 to 1348. Revolution 1348 closed the seasonal visibility window of \inte\ in the direction of the source
and no other data were collected after 56594~MJD (see Table~\ref{tab:spec}).  
The source was generally observed at relatively high off-axis angles and was included in the FOV of the two JEM-X units 
\citep{lund03} for only a relatively small fraction of the total observational time (revolutions 1346 to 1348). 
The analysis of INTEGRAL data was carried out by  the OSA10.0 distributed 
by the ISDC \citep{courvoisier2003}. We show in Fig.~\ref{fig:ima} the ISGRI mosaic of the  field of view (FOV) around the source. 
In this mosaic (total effective exposure time 137~ks) the source is detected with a significance of 41$\sigma$ in the 20-100~keV energy band. 
In the JEM-X mosaic (effective exposure time 6~ks), the source was detected at a significance 
of 13$\sigma$ in the the 3-35 keV energy band. To perform a broad-band spectral analysis of the X-ray emission recorded from the 
source, we extracted the ISGRI and JEM-X spectra during time intervals quasi-simultaneous with \swift\ data (see Table~\ref{tab:spec}). 

\section{\swift\ data}

\swift\ observations were carried out immediately after the discovery of the source and covered about 30 days of the 
outburst \citep{kennea2013,kennea2013a}. The first two pointings (ID~00032993001, 00032994001) were collected in photon-counting mode (PC; time resolution 2.5~s) and  were
severely affected by pile-up due to the high flux of the source \citep{kennea2013}. We therefore discarded these data for the following analysis. 
All the remaining data were collected in window-timing mode (WT; time resolution 1.8~ms, see Table~\ref{tab:swift}). 
XRT data were processed with the {\sc xrtpipeline} (v.0.12.6) and were analyzed using standard procedures 
\citep{burrows2005}. We selected only event grades 0-2 and adopted the latest response files available (v.014).  
Exposure maps were created through the {\sc xrtexpomap} task and were used to generate ARF files. The source light curves in each 
observations were corrected for all instrumental issues with the task {\sc xrtlccorr}. 
All source event lists were barycentered by using the {\sc barycorr} tool.  
In all XRT observations the source was detected at count-rates 150-250~cts/s in energy band 0.3-10~keV,  depending also on the position of the source on the detector and the presence of bad columns\footnote{http://www.swift.ac.uk/analysis/xrt/exposuremaps.php}. 
We followed the technique described in \citet{romano2006} to search
for pile-up in the XRT observations, and found that excluding the two innermost pixels of the extraction region centered on the  
source was sufficient to correct for this problem in all pointings. Correction for pile-up 
was not needed for pointing 8, as in this case a bad column passing through the source already decreased the count-rate below the level at which pile-up is a problem. 
All XRT spectra were grouped to have at least 100 photons per energy bin and were fit with the $\chi^2$-statistic in {\sc Xspec}.  
Uncertainties throughout the paper are given at 90\% c.l. if not stated otherwise.

\section{Results}
\label{sec:results}

\begin{figure}
\centering{
\includegraphics[scale=0.4]{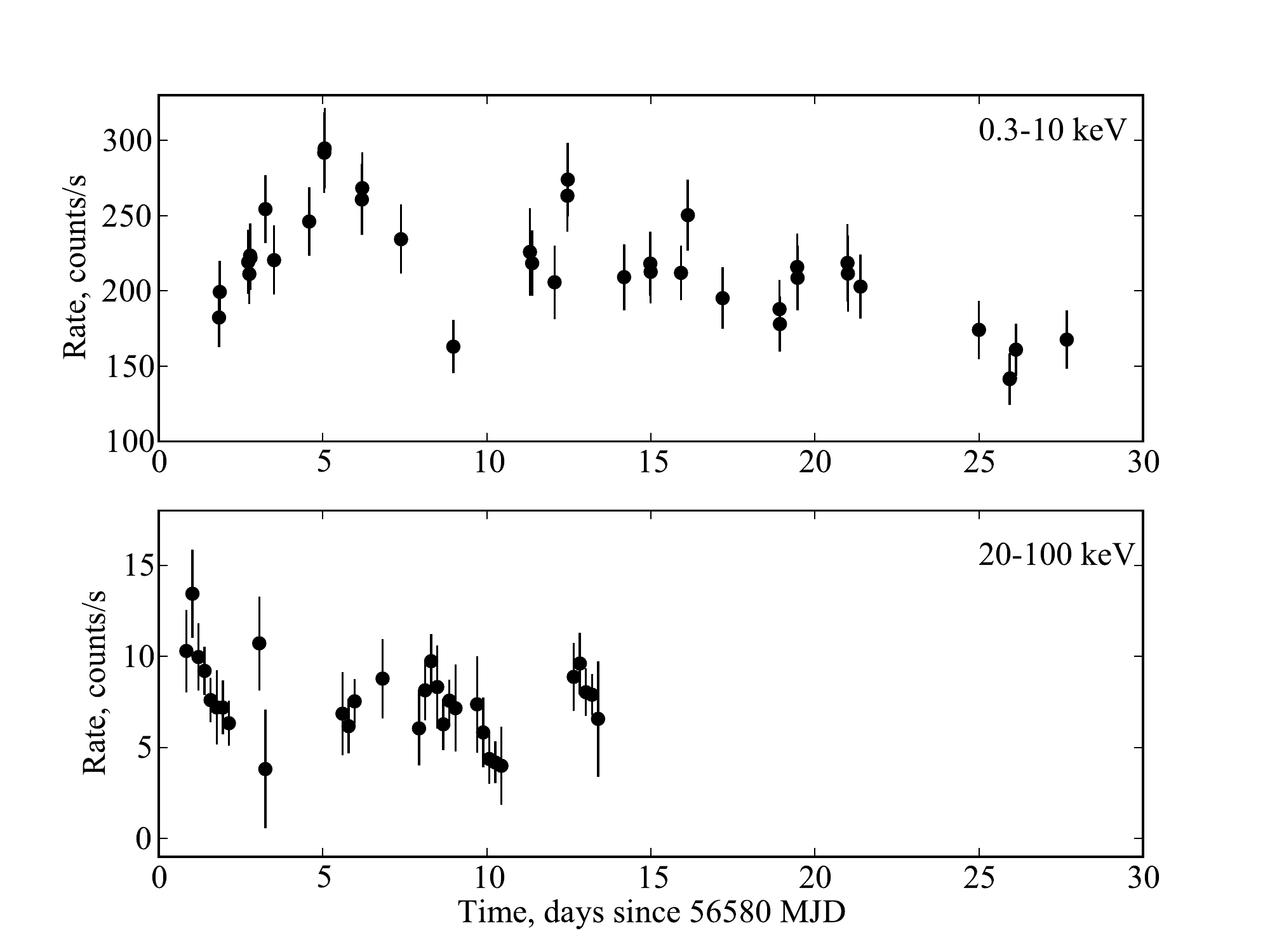}
\caption{ \maxi\ lightcurve during the outburst as observed by \swift\,/XRT (0.3-10 keV, upper panel) and \inte\,/ISGRI (20-100 keV, lower panel).  
 The time bin is 1~ks for XRT and 16~ks for ISGRI.}
}
\label{lc}
\label{fig:curve} 
\end{figure}

The XRT and ISGRI light curves of the source covering the entire observational campaign reported in this paper are shown in 
Fig.~\ref{fig:curve}. The XRT light curve shows a noticeable increase of the source X-ray flux (0.3-10~keV) during the first five days of the outburst with a gradual decrease afterwards. In the ISGRI energy band (20-100~keV) no significant flux evolution is detected (though with large uncertainties). 
This behavior agrees with that found in the publicly available MAXI data of the 
source\footnote{http://maxi.riken.jp/top/index.php?cid=1\&jname=J1828-250}.

\begin{figure*}
\centering
\includegraphics[angle=-90, width=2\columnwidth]{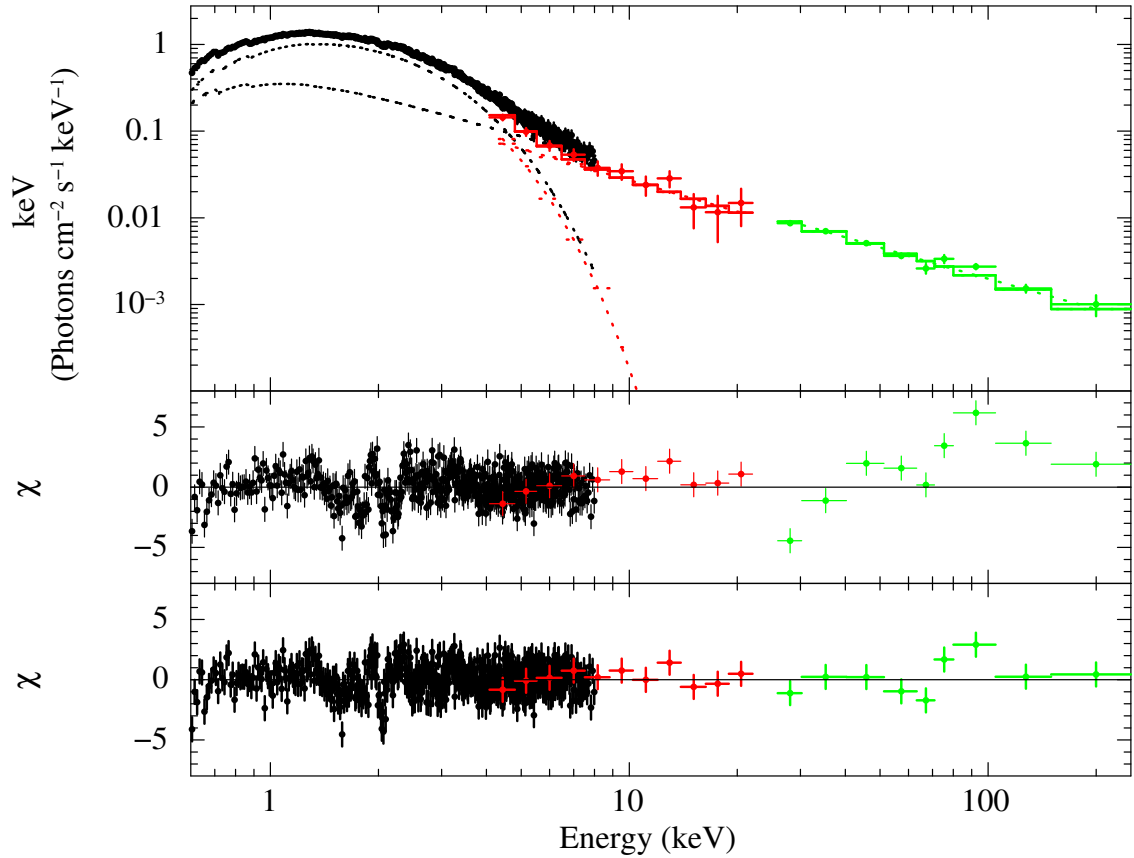}
\vspace{-0.5truecm}
\caption{Averaged \maxi\, broad-band spectrum obtained by integrating over all available \swift/XRT (black), JEM-X (red) and ISGRI (green) data from MJD 56580 to MJD 56594 ("total" spectrum in Table 1). The best-fit model is shown with a solid line. 
The two dashed lines represent the spectral components (the multi-temperature disk black-body and the power-law). Residuals from the best 
fit are reported in the middle (no gain correction applied) and bottom panel (after gain correction). Some instrumental residuals are 
still visible at energies 1.5-2.5~keV also after the gain correction but their significance is strongly reduced.}
 \label{spec}
\label{fig:spec}       
\end{figure*}

We first extracted all available XRT spectra of the source and fit them with a simple absorbed power-law model. In all cases 
these fits gave unacceptable results ($\chi^2_{\rm red}$$\gg$1) and showed a clear excesses in the residuals at energies $\lesssim$2-3~keV. 
Data points at energies $<$0.6~keV were discarded from the analysis to avoid strong instrumental residuals 
around the O-K edge. 
We found that a model comprising an absorbed multi-temperature disk black body ({\sc diskbb} in {\sc Xspec}) and a power law could 
be used to describe all XRT spectra in the energy range 0.6-10~keV equally well. In all cases we obtained 
an absorption column density consistent with the Galactic value expected in the direction of the 
source \citep[$\sim$0.3$\times$10$^{22}$~cm$^{-2}$,][]{dickey1990}, an inner disk radius (temperature) of 
$R_{bb}\sim$25~km ($kT_{bb}\sim$0.7~keV) and a power-law photon index of $\Gamma \sim$3. 
When the INTEGRAL data were also used to perform a broad-band fit, the higher energy emission of the source was characterized by a significantly harder power-law spectrum ($\Gamma$=2.2) than that derived from the XRT 
spectra. We found that this discrepancy arises from instrumental residuals in the XRT spectra at energies 
1.5-2.5~keV, corresponding to the $\mathrm{Au-M_V}$ and $\mathrm{Si-K}$
edges\footnote{http://www.swift.ac.uk/analysis/xrt/files/SWIFT-XRT-CALDB-09\_v17.pdf}. 
These residuals are tackled by the $\chi^2$ minimization routine in {\sc Xspec} mainly by adjusting the 
power-law photon index (poorly determined in the fits to the XRT spectra alone because of the limited energy coverage of the instrument).  
The instrumental residuals in XRT become very evident when all data are summed to extract an average spectrum (see Fig.~\ref{fig:spec}). 
Following the suggestion in the latest available XRT calibration document, we tried to account for these residuals in the fits to 
the XRT data alone by fixing $\Gamma=2.2$ (as determined from the average broad-band spectrum) and using the {\sc gain} correction 
in {\sc Xspec}. The results of the broad-band spectral analysis obtained with this method are summarized in 
Table~\ref{tab:spec} (normalization constants were introduced in all broad-band fits to account for the variability of 
the source and intercalibration between the different instruments). We indicated in all cases the measured correction to the 
energy slope (Sl) and shift (G) of the XRT spectra. For consistency, the gain correction was also applied during the 
fits to the XRT spectra alone, as reported in Table~\ref{tab:swift} (in several cases the values of Sl and G were poorly
constrained and did not significantly affect the fits because of the relatively low statistics of the data). 
The hard X-ray emission from the source is clearly detected up to $\sim$200~keV. 

\begin{table*}
\caption{Results of the broad-band spectral analysis.}              
\label{tab:spec}       
\centering  
\scriptsize
\begin{tabular}{@{}ccccccccccccccccl@{}}    
\hline\hline                        
Data\tablefootmark{1} & \multicolumn{3}{c} {Exposures} & $N_H$$^2$& $\Gamma$& $kT_{bb}$&$R_{disk}$$^3$ & $C_{JEM-X}$&$C_{ISGRI}$ &\multicolumn{2}{c}{Flux$^4$}& Sl$^5$ & G$^5$ & $\chi^2/d.o.f.$\\   
& \multicolumn{3}{c} {(ks)} &&& keV& (km)&&&0.6-10 &20-100 & & eV \\ 
&XRT&JEM-X&ISGRI &&&&&&&keV&kev&\\
\hline                                   
1344+01&1.0&-&26&0.19$\pm$0.02 &2.0$\pm$0.16&0.68$\pm$0.02&  28.4$\pm$1.8&-&0.6$\pm$0.2& 4.9&0.6&0.96$\pm$0.01&37$\pm$25&1.2/283\\
1344+03&1.0&-&8 &0.32$\pm$0.08&2.68$\pm$0.24&0.73$\pm0.03$&22.7$\pm$2.5&-&1.9$\pm$0.9&6.1&0.7&0.99$\pm$0.02&0$\pm$30&1.2/331\\
1345+06&1.0&-&30&0.2$\pm$0.02&2.1$\pm$0.2&0.76$\pm0.01$&25.3$\pm$1.7&-&$0.7\pm0.2$&6.1&0.6&0.97$\pm$0.01&35$\pm$18&1.2/338\\
1346+07&0.9&-&9&0.18$\pm$0.03&1.86$\pm$0.24&0.74$\pm0.02$&27.8$\pm$1.7&-&$0.5\substack{+0.2 \\ -0.1}$&6.9&0.8&0.92$\pm$0.02&19$\substack{+32 \\ -27}$&1.0/310\\
1346+08&0.2&1.7&17&0.21$\pm$0.04 &2.1$\pm$0.2&0.7$\pm$0.04& 22.7$\pm$3.3&$2.3\substack{+0.9 \\ -0.7}$&$1.4\substack{+0.8 \\ -0.4}$& 3.5&0.7&1.03$\pm$0.03&25$\substack{+30 \\ -50}$&1.0/121\\
1347+11&0.9&3.4&30&0.25$\pm$0.05 &2.3$\pm$0.3 &0.68$\pm$0.02& 29$\pm$3&$0.5\pm0.2$&$0.5\substack{+0.4 \\ -0.2}$& 5.7&0.4&0.98$\pm$0.02&15$\pm$20&1.2/251\\
1348+12&1.05&3.4&40&0.2$\pm$0.02 &2.2$\pm$0.2 &0.72$\pm$0.01& 25.6$\pm$1.4 &0.7$\pm$0.1&0.63$\pm$0.03& 6.1&0.6& 0.95$\pm$0.02 & 45$^{+20}_{-22}$ &1.1/211\\
Total & 10.5 & 8.6 & 160 & 0.22$\pm$0.01 & 2.2$\pm$0.1 &0.72$\pm$0.04& 25.4$\pm$0.5 &0.7$\pm$0.1&0.8$\pm$0.1& 5.9&0.6& 0.97$\pm$0.03 & 24$\pm$6 &1.4/614\\
\hline                                             
\end{tabular}\\
\vspace{-0.2truecm}
\tablefoot{$^1$Indicates the INTEGRAL revolution + the latest two digits XX of the Swift observation ID~000329970XX. 
$^2$In units of $\times 10^{22} cm^{-2}$. $^3$The radius of 
the disk is calculated assuming $\cos{\theta}$=0 and a distance to the system of 8\,kpc. 
$^4$In units of $\times 10^{-9}$ erg s$^{-1}$ cm$^{-2}$. 
$^5$G (Sl) is the energy shift (slope) of the gain correction derived from the spectral fit in {Xspec}.} 
\end{table*}

\begin{figure}
\centering
\includegraphics[angle=0,scale=0.3]{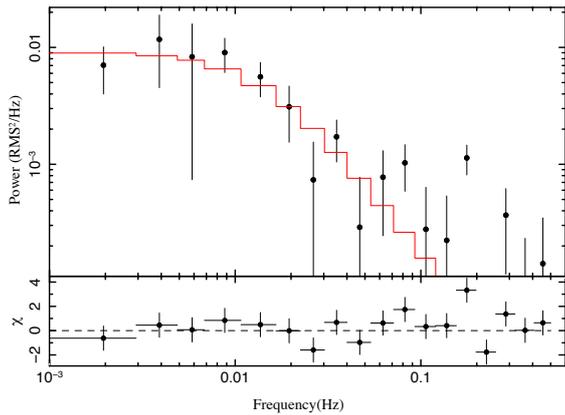}
\vspace{-0.5truecm}
\caption{Averaged PSD of MAXI J1828-249 obtained from all available XRT data. 
The PSD here is rebinned with a logarithmic step of 1.25. The white noise contribution has been subtracted. 
The red solid line represents the best fit obtained with a zero-centered Lorentzian (see text for details).}
\label{fig:psd}
\end{figure}

\begin{figure}
\centering
\includegraphics[width=\columnwidth,angle=0]{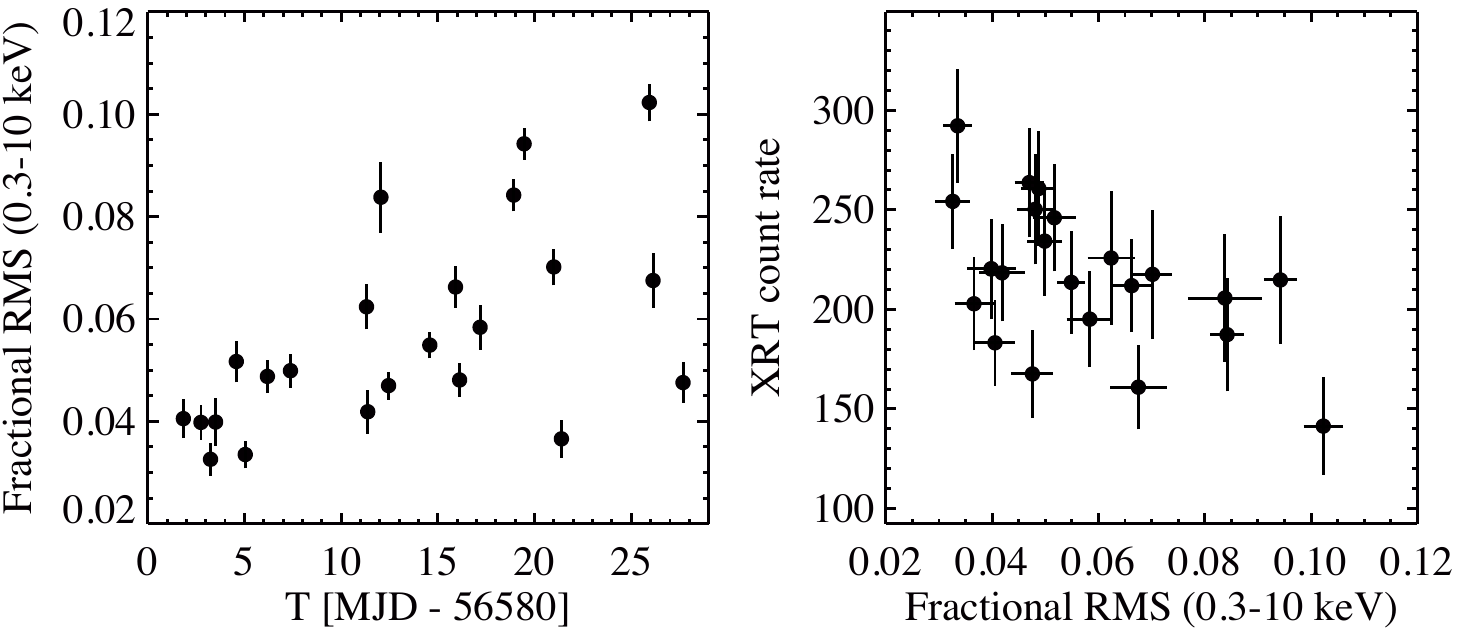}
\vspace{-0.5truecm}
\caption{Fractional rms of \maxi\ (4-50\,mHz frequency range) measured from the XRT data. The left panel shows 
the rms as function of time; in the right panel the average XRT count-rate of each pointing is plotted as a function of 
the fractional rms.}
\label{fig:rms}
\end{figure}

The timing analysis of the \swift\,/XRT data was carried out with a technique similar to that reported in \citet{ferrigno2012}. 
We first searched for timing features in the event files extracted from each of the XRT pointing, but no 
statistically significant features were found. We therefore computed the average power spectral density 
(PSD) of the source by performing a Fourier transform on its light curves (time bin of 1\,s) 
and averaging on intervals of 512\,s 
(corresponding to roughly 1-2 time intervals per observation). The PSD were well fit 
($\chi^2_\mathrm{red}/\mathrm{d.o.f.}=1.3/33$) by a zero-centered Lorentzian plus white noise. The latter was added as a constant $C$ 
to the fit and we measured in Leahy normalization $C$=$2.17\pm0.07$. The source PSD is shown in Fig.~\ref{fig:psd}.
We estimated the rms variability of the source from the zero-centered Lorentzian after subtracting the contribution of the white noise 
and obtained rms=0.90$_{-0.13}^{+0.11}$\% and $\nu_0=14^{+8}_{-4}$\,mHz. Because this average analysis showed that the aperiodic noise 
is more prominent in the 4-50\,mHz frequency range, we used this interval to compute the fractional rms of each observation separately. 
The results are reported in Fig.~\ref{fig:rms}. The uncertainties on the rms of each observation was obtained from bootstrapping simulations 
with 1000 realizations \citep[for a detailed description of the method see][]{ferrigno2012}. 
The source exhibited an rms $\lesssim$10\%  in all XRT pointings and no particular trend is 
found when the rms is plotted as a function of the total source count-rate (see Fig.~\ref{fig:rms}).
 
Owing to the relatively low statistics of the \inte\ data, we did not attempt a timing analysis of the hard X-ray 
emission from \maxi.\  

\begin{table*}
\caption{ \swift\ data spectral fits.}              
\label{tab:swift}       
\centering  
\scriptsize                                    
\begin{tabular}{@{}c c c c c c c c c c c c c c @{}}    
\hline\hline                        
ObsID$^1$ &Start Time& Stop Time & Exposure& $N_H$& Photon index$^3$& $kT_{bb}$&$R_{disk}$$^2$ &Flux (0.6 -- 10 keV) & Sl$^5$ & G$^5$ & $\chi^2$/d.o.f\\   
&MJD& MJD & ks&($\times 10^{22} cm^{-2}$)&$\Gamma$& keV& km&$10^{-9}$ erg s$^{-1}$ cm$^{-2}$& & eV & \\ 
\hline
1   &56581.831 & 56581.860 &0.97 & 0.21$\pm$0.02 & 2.2 & 0.67$\pm$0.02 & 28.1$^{+1.6}_{-1.4}$ & 4.9 & 0.96$\pm$0.01 & 29$^{+26}_{-20}$ & 1.1/275\\
2   &56582.720 & 56582.791 &0.98 & 0.21$\pm$0.03 & 2.2 & 0.72$\pm$0.02 & 26.4$^{+1.7}_{-1.2}$& 6.4 & 0.96$\pm$0.01 & 45$^{+16}_{-28}$ & 1.0/311\\
3   &56583.244 & 56583.256 &0.95 & 0.22$\pm$0.02 & 2.2 & 0.71$\pm$0.01 & 26.5$\pm$1.2 & 6.2 & 0.97$\pm$0.01 & 20$^{+20}_{-21}$& 1.2/324\\
4$^4$   &56583.509 & 56583.517 &0.68 & 0.23$\pm$0.03 & 2.2 & 0.70$\pm$0.03 & 28.2$^{+3.4}_{-3.2}$ & 6.5 & 0.97$\pm$0.02 & 12$^{+45}_{-43}$ & 1.1/250\\
4$^4$   &56584.579 & 56584.586 &0.59 & 0.23$\pm$0.03 & 2.2 & 0.73$\pm$0.02 & 29.6$^{+2.1}_{-2.3}$ & 6.8 & 0.97$\pm$0.02 & 7$^{+35}_{-27}$ & 1.1/265\\
5   &56585.039 & 56585.053 &1.20 & 0.21$\pm$0.02 & 2.2 & 0.75$\pm$0.01 & 27.8$^{+1.9}_{-1.3}$ & 6.8 & 1.01$\pm$0.01& 0$^{+18}_{-28}$ & 1.2/352\\
6   &56586.185 & 56586.196 &1.01 & 0.20$\pm$0.01 & 2.2 & 0.78$\pm$0.01 & 24.9$^{+1.1}_{-1.0}$ & 6.1 & 0.97$\pm$0.01& 35$\pm$16 & 1.1/330\\
7   &56587.378 & 56587.389 &0.94 & 0.22$\pm$0.02 & 2.2 & 0.74$\pm$0.02 & 27.3$\pm$2.0 & 6.7 & 0.98$\pm$0.02 & 0$^{+32}_{-17}$ & 1.0/303\\
8   &56588.973 & 56588.975 &0.18 & 0.22$\pm$0.03 & 2.2   & 0.70$\pm$0.04 &23.0$\pm$3.0 & 3.5 & 1.03$\pm$0.04 & 0$^{+35}_{-43}$ & 0.97/107\\
11$^4$ &56591.308 & 56591.318 &0.90 &0.23$\pm$0.02 & 2.2 & 0.68$\pm$0.02 & 30.0$\pm$2.0 & 5.7 & 0.98$\pm$0.02 & 17$\pm$31 & 1.2/239\\
11$^4$ &56591.375 & 56591.382 &0.61 &0.23$\pm$0.02 & 2.2 & 0.71$\pm$0.03 & 28.1$\pm$2.1 & 6.1 & 0.98$\pm$0.02 & 27$^{+25}_{-46}$ & 1.1/253\\
12 &56592.450 & 56592.462 &1.05 &0.21$\pm$0.01 & 2.2 & 0.71$\pm$0.02 & 26.2$^{+1.8}_{-1.0}$ & 6.0 & 0.95$\pm$0.01 & 39$\pm$20 & 1.2/324 \\
13 &56592.058 & 56592.062 &0.34 &0.22$\pm$0.05 & 2.2 & 0.70$\pm$0.04 & 31.7$\pm$6.3 & 3.6 & 1.02$\pm$0.02 & 0$^{+56}_{-72}$ & 0.98/164 \\
14 &56594.177 & 56594.994 &1.48& 0.22$\pm$0.02 & 2.2 & 0.72$\pm$0.02 & 28.9$\pm$1.5 & 5.9 & 0.95$\pm$0.01 & 33$^{+22}_{-27}$ & 1.2/330\\
15 &56595.912 & 56595.921 &0.74 &0.24$\pm$0.03 & 2.2 & 0.70$\pm$0.02 & 30.2$^{+3.1}_{-2.2}$ & 6.2 & 0.98$\pm$0.01 & 0$^{+30}_{-40}$ & 1.1/302\\
16 &56596.120 & 56596.132 &0.98 &0.20$\pm$0.01 & 2.2 & 0.71$\pm$0.01 & 27.3$\pm$1.3 & 5.6 & 0.95$\pm$0.01 & 40$^{+16}_{-23}$& 1.1/308\\
17 &56597.180 & 56597.189 &0.76 &0.22$\pm$0.01 & 2.2 & 0.67$\pm$0.02 & 28.6$^{+2.8}_{-1.7}$ & 5.8 & 0.95$\pm$0.01 & 40$^{+18}_{-21}$ & 1.1/258\\
18 &56598.915 & 56598.927 &1.05 &0.24$\pm$0.04 & 2.2 & 0.68$\pm$0.03 & 28.4$^{+4.0}_{-2.0}$ & 4.6 & 1.01$\pm$0.01& 0$^{+32}_{-50}$ & 0.98/294\\
19 &56599.453 & 56599.467 &1.16 &0.20$\pm$0.01 & 2.2 & 0.71$\pm$0.02 & 25.5$^{+3.7}_{-1.4}$ & 4.9 & 0.95$\pm$0.02 & 51$^{+26}_{-51}$& 1.2/299\\
20 &56600.987 & 56601.000 &1.15 &0.21$\pm$0.01 & 2.2 & 0.68$\pm$0.02 & 24.9$\pm$1.7 & 4.9 & 0.95$\pm$0.01 & 38$^{+18}_{-22}$ & 1.1/273\\
21 &56601.384 & 56601.393 &0.80 &0.21$\pm$0.02 & 2.2 & 0.67$\pm$0.02 & 25.3$\pm$1.6 & 4.7 & 0.96$\pm$0.01& 33$^{+18}_{-22}$& 1.1/264\\
22 &56604.994 & 56604.997 &0.19 &0.3$\pm$0.1 & 2.2 & 0.6$\pm$0.1 & 38.5$^{+12.3}_{-15.3}$ & 4.8 & 1.0$\pm$0.1 & 0$^{+194}_{-93}$ & 1.1/108\\
23 &56605.928 & 56605.940 &1.05 &0.21$\pm$0.01 & 2.2 & 0.62$\pm$0.02 & 26.9$\pm$2.2 & 4.6 & 0.95$\pm$0.01 & 41$^{+22}_{-18}$ & 1.3/258\\
24 &56606.121 & 56606.130 &0.75 &0.25$\pm$0.03 & 2.2 & 0.61$\pm$0.02 & 29.4$\pm$2.8 & 4.6 & 0.96$\pm$0.02 & 12$^{+33}_{-27}$ & 1.0/251\\
25 &56607.670 & 56607.681 &0.95 &0.22$\pm$0.02 & 2.2 & 0.64$\pm$0.02 & 31.1$^{+3.2}_{-1.6}$ & 5.7 & 0.96$\pm$0.01 & 36$^{+18}_{-29}$ & 1.2/253\\

\hline                                             
\end{tabular}
\vspace{-0.2truecm}
\tablefoot{$^1$XRT observation ID as in Table~\ref{tab:spec}. $^2$The radius of the disk is 
estimated as described in the notes of Table~\ref{tab:spec}. $^3$The value of $\Gamma$ was fixed to that of the 
averaged broad-band spectrum (Fig.~\ref{fig:spec}). $^4$The observation was divided into two snapshots that we 
analyzed separately. $^5$As in Table~\ref{tab:spec}.}
\end{table*}

\section{Discussion}
\label{sec:discussion}

\maxi\ was discovered in outburst on 2013 October 15 by the MAXI/GSC instrument and apparently underwent a 
spectral transition from a hard to a softer state about one day later \citep{negoro2013}. Such transitions 
are typically observed in the so-called black hole candidates (BHC), which suggests that \maxi\ is associated 
with this class of objects.
According to the standard scenario \citep{fender2004,homan2005,belloni2010}, BHC sources are known to evolve during their 
outburst along a q-shaped track in the hardness-intensity diagram. The outburst starts in the low-hard spectral state (LHS), which is 
characterized by a power-law shaped X-ray spectrum with $\Gamma\sim1.6-1.7$ and a cut-off at the higher energies $E_{cut} \sim 100$ keV. 
Radio emission arises in this state due to the synchrotron radiation of a steady jet. In the following phases of the outburst,  
the X-ray and radio luminosities both increase until the source reaches the high-soft state (HSS), characterized by a prominent thermal emission 
from the accretion disk and a marginal power-law tail. Radio emission at this state is no longer observed, most likely because of the suppression of the jet. 
The time variability of the source is also significantly different in the two states, being generally more pronounced in the LHS with rms values of up to 
$\sim$30\% and characterized by the quasi-periodical oscillations (QPOs) in the HSS in combination with a strongly suppressed rms \citep[see, e.g.,][for a recent review]{belloni2010}. In addition to these two main states, at least two intermediate hard and soft states were identified. 
In particular, BHCs in the so-called intermediate soft state still show a significant hard component (extending to hundreds of keV) though they are no longer significantly detected in the radio domain and present both a prominent soft spectral component and a limited rms. 
 Not all transient BHCs in outburst go through a complete q-track. So far, a limited number of objects were observed that reach the hard intermediate state during outburst, but then return to the LHS instead of moving to HSS \citep[see, e.g.,][]{capitanio2009,ferrigno2012}.

 The observational campaign we presented in this paper started about 1.8~days after 
the discovery, that is, immediately after the possible spectral transition reported by \cite{negoro2013}. 
The broad-band energy spectrum of the source 
as measured by  \swift\ and \inte\ comprised a thermal component that we associated with the emission from a 
multi-temperature black-body disk, and a hard power-law extending with no measurable break up to 200~keV. 
Assuming a distance to the source of 8~kpc, its peak luminosity is $\sim10^{38}$ erg/s. 
During the entire observational campaign \maxi\ did not show significant spectral evolution. 
Our observational results suggest that \maxi\, is a transient BHC that evolved rapidly in the first two days of the outburst 
from the LHS to the soft intermediate state and remained in this state during the entire observational campaign presented in this work.    
This conclusion is supported by both the spectral and timing analysis\footnote{Note that the range of 
frequencies we could investigate in the case of \maxi\ based on the available X-ray data is slightly lower (0.001-0.5 Hz) compared to that 
in which typical QPOs for this state are observed (0.01 - 50 Hz). However, the measured rms level in the XRT data is fully compatible with that  
commonly found in the soft intermediate state.} reported in Sect.~\ref{sec:results} and the lack of significant radio emission 
already two days after the onset of the outburst, as mentioned in Sect.~\ref{sec:intro}.
The light curve of the source based on MAXI publicly available data also supports this scenario and suggests that the source is now fading into quiescence. This means that \maxi\, might represent another example of transient BHC ''failed'' outburst,
as mentioned above. Future observations in mid-2014 (e.g. with {\it Swift}/XRT) 
will be able to confirm/reject this conclusion.

{\bf Note added to proofs.}
After this paper was accepted for publication, new \swift\, data
collected on 2014 February 14 found that the source is back in a
faint hard state (power-law photon index $\sim$1.7 and 0.5-10 keV flux of
$4\times 10^{-11}$ erg/cm$^2$/s; \cite{tomsick2014}). Radio
observations performed on 2014 February 16 also detected significant
radio emission from the source (preliminary flux densities 1.38$\pm$0.05 mJy at 5.5 GHz and 1.28$\pm$0.06 mJy at 9 GHz; \cite{corbel2014}), which supports our conclusions in Sect. 5.

\section*{Acknowledgments}
EF acknowledges support by grant NSh-6137.2014.2, RFBR 13-02-00741 and grant 20FI20\_135263 
of the National Swiss Fond.

\bibliographystyle{aa}
\bibliography{maxij1828.bib}

\end{document}